\documentclass[12pt,english]{article}
\usepackage{times}
\usepackage[T1]{fontenc}
\usepackage[latin1]{inputenc}
\usepackage{graphicx}
\usepackage{geometry}
\usepackage{setspace}
\usepackage{amssymb}

\makeatletter

\newcommand{\boldsymbol}[1]{\mbox{\boldmath $#1$}}

\usepackage{setspace}

\usepackage{babel}
\makeatother
\begin{document}

\title{\textbf{\large Laboratory experiments on dynamo action and 
   magnetically triggered flow instabilities}}

\renewcommand{\thefootnote}{\fnsymbol{footnote}}
\author{F. STEFANI\footnote{Corresponding author. Email: F.Stefani@hzdr.de} $\dagger$, 
A. GAILITIS$\ddagger$,
G. GERBETH$\dagger$, 
A. GIESECKE$\dagger$,\\
Th. GUNDRUM$\dagger$,
G. R\"UDIGER$\S$,
M. SEILMAYER$\dagger$,
T. VOGT$\dagger$}

\maketitle

\begin{center}
$\dagger${\small Helmholtz-Zentrum Dresden-Rossendorf, Bautzner 
Landstr. 400, D-01328 Dresden, Germany}\\
$\ddagger${\small Institute of Physics, Latvian University, LV-2169 Salaspils 1, Riga, Latvia}\\
$\S${\small Leibniz-Institut f\"ur Astrophysik Potsdam, An der Sternwarte 16, D-14467 Potsdam, Germany}
\end{center}

{\footnotesize Magnetic fields of planets, stars and galaxies are generated 
by self-excitation in moving electrically conducting fluids. Once produced, 
magnetic fields can play an active role in cosmic structure formation 
by destabilizing rotational flows that would be otherwise hydrodynamically 
stable. For a long time, both hydromagnetic dynamo action as well as 
magnetically triggered flow instabilities had been the subject of purely 
theoretical research. Meanwhile, however, the dynamo effect has been
observed in large-scale liquid sodium experiments in Riga, 
Karlsruhe and Cadarache. In this paper, we 
summarize the 
results of 
liquid metal experiments devoted to the dynamo effect and
various magnetic instabilities such as the 
helical and the azimuthal magnetorotational instability and  
the Tayler instability. 
We discuss in detail our plans for a  
precession-driven dynamo experiment and a large-scale
Tayler-Couette experiment using liquid sodium, and
on the prospects to observe magnetically triggered 
instabilities of flows with positive shear.
}

\noindent \textit{\footnotesize Keywords:} {\footnotesize Dynamo;
Instabilities
}{\footnotesize \par}

\section{Introduction}

Self-excitation in moving electrically conducting fluids, 
i.e. hydromagnetic dynamo action, is well-known to be responsible 
for the generation of planetary, stellar and galactic magnetic 
fields (R\"udiger {\it et al.} 2013). Yet, cosmic magnetic fields, 
rather than being only passive by-products of fluid motion, play 
also an active role in the formation of stars and black-holes 
by fostering angular momentum transport and mass accretion 
via the magnetorotational instability (MRI) (Balbus 2003). 
Complementary to theoretical and numerical 
modeling of these fundamental problems of geo- and astrophysical 
magnetohydrodynamics (MHD), the last two decades have seen 
great progress in respective experimental studies (Gailitis {\it et al.} 2002,
Stefani {\it et al.} 2008, Lathrop and Forest 2011). 
By now, the "magic" threshold for magnetic field 
self-excitation has been attained in three large-scale 
liquid sodium experiments in Riga 
(Gailitis {\it et al.} 2000, 2003, 2004), Karlsruhe 
(M\"uller and Stieglitz 2000, Stieglitz and M\"uller,
2001, M\"uller  {\it et al.} 2004), and Cadarache 
(Monchaux  {\it et al.} 2007,
Monchaux  {\it et al.} 2009, Miralles  {\it et al.} 2013). Dynamo and MRI related liquid metal 
experiments were carried out, or are under construction, all over 
the world, e.g. in Madison 
(Spence {\it et al.} 2006), Maryland (Zimmerman 
{\it et al.} 2010), Socorro (Colgate {\it et al.} 2011), 
Princeton (Nornberg {\it et al.} 2010), Perm (Frick {\it et al.} 2010), 
Grenoble (Schmitt {\it et al.} 2013), and Zürich (Hollerbach {\it et al.} 
2013). An MRI-like coherent structure of velocity and magnetic fields 
had been observed in a liquid sodium spherical Couette flow 
at the University of Maryland (Sisan {\it et al.} 2004), 
albeit on the background 
of an already  turbulent flow. Both the helical (HMRI) and 
the azimuthal (AMRI) versions of MRI were observed and explored 
in much detail in the experiment PROMISE at Helmholtz-Zentrum 
Dresden-Rossendorf (HZDR) (Stefani {\it et al.} 2006, Stefani {\it et al.} 
2009,
Seilmayer {\it et al.} 2014). The current-driven, 
kink-type Tayler instability (TI) (Tayler 1973), 
which is at the root 
of an alternative model of the solar dynamo (Spruit 2002), was 
demonstrated in a long column of a liquid 
metal, just by running a strong electrical current through 
it (Seilmayer {\it et al.} 2012). 

More often than not, 
the need to design and optimize 
those experiments, and to understand their occasionally 
unexpected results, has incited complementary theoretical 
and numerical work, which in some cases also spurred new 
discussions in the original geo- or astrophysical domain. 
Moreover, the experiments have triggered the development of 
flow measuring techniques also applicable to steel casting and 
crystal growth, e.g. Contactless Inductive Flow 
Tomography (CIFT) (Wondrak {\it et al.} 2010), 
and of various stabilization 
methods for large-scale liquid metal batteries 
(Stefani {\it et al.} 2016a).

The first objective of this paper is to provide a cursory,
incomplete, and personally biased
account of those previous experiments that were 
devoted to dynamo action and magnetically triggered 
flow instabilities. 
Then, we will discuss in more detail the plans for three 
new liquid sodium experiments. 
These comprise, first, a  
truly homogeneous dynamo in form of a large-scale 
precession experiment. Besides thermal and compositional 
buoyancy, precession has long been discussed as an 
alternative, or at least additional, power 
source of the geodynamo (Malkus 1968,
Kerswell 1993, Tilgner 2005, Tilgner 2007, 
Wu and Roberts 2009, Shalimov 2006), the ancient 
lunar dynamo (Dwyer {\it et al.} 2011, Noir and Cebron 2013, 
Weiss and Tikoo 2014), 
or the asteroid Vesta (Fu {\it et al.} 2002). 
The second experiment will deal 
with different versions of the MRI, and their combinations 
with the TI, with the main goal to find the "holy grail" 
of this research field, which is standard MRI (SMRI) 
with a purely 
axial field being applied 
(Velikhov 1959, Balbus and Hawley 1991). 
Third, we will discuss a possible experimental 
demonstration of  the recent prediction that 
rotational flows with a radially increasing angular velocity, 
such as it prevails in an equator-near-strip of the solar 
tachocline, might be destabilized by purely azimuthal 
magnetic fields (Stefani and Kirillov 2015, 
R\"udiger {\it et al.} 2016, R\"udiger {\it et al.} 2018). 

All these experiments will be 
carried out in frame of the liquid sodium 
platform DRESDYN (DREsden Sodium facility for DYNamo and 
thermohydraulic studies) 
at HZDR (Stefani {\it et al.} 2012, Stefani {\it et al.} 2015), 
whose building construction has been finalized in 
2017.

\section{Previous experiments}

In this section, we briefly summarize the 
previous experiments on 
dynamo action and various magnetically triggered
flow instabilities. For more details, see the survey papers
by Gailitis {\it et al.} (2002), Stefani {\it et al.} (2008), and
Lathrop and Forest (2011).

\subsection{Dynamo (related) experiments}

For many decades, it had been a dream to simulate a homogenous, 
near-natural dynamo in the laboratory. In the 
1960ies, based on Herzenberg's theoretical derivation of a 
working dynamo (Herzenberg 1958), Lowes and Wilkinson (1963,1967)
conducted a
series of quasi-homogeneous dynamo experiments consisting of 
two or more rotating soft-iron cylinders embedded into 
blocks of the same material. 
These experiments suffered from the hysteresis 
effect of the used magnetic material, and from the 
rigidity of their construction which made the saturation 
mechanism very unlike to that of a fluid dynamo. However, their 
step-by-step improvement even led to a sort of reversals 
similar to those of the geomagnetic field (Wilkinson 1984), 
for whose explanation the oscillatory solutions of
Herzenberg's dynamo, as first found by Gailitis (1973),
might play a role.  

Proposed by Steenbeck, a liquid sodium experiment 
with two orthogonally 
interlaced copper channels was carried out in 1967. 
Reaching flow velocities of up to 11 m/s, this 
experiment provided clear evidence of 
the $\alpha$-effect, i.e. of the  induction of an 
electromotive force 
{\it parallel} to the applied magnetic field (Steenbeck 
{\it et al.} 1967).
Due to its relevance in our context, we 
mention the 
precession experiment with liquid sodium 
reported by Gans (1971). 
The sodium was enclosed in a cylindrical 
volume of 25 cm diameter and approximately the 
same height. Rather high rotation rates of up 
to 60 Hz were attained, but the precession rate 
was quite low (below 1 Hz). Nevertheless, Gans was 
able to measure a magnetic field amplification by a 
factor of 3. A wealth of knowledge on precession 
driven flows has been gained with 
water experiments  
in various geometries
(Malkus 1967, Vanyo and Dunn 2002, Lagrange {\it et al.} 2008,
Mouhali {\it et al.} 2012, Goto {\it et al.} 2014, 
Lin {\it et al.} 2015 ). 
With the cylindrical water experiment at HZDR
(see Figure 4f further below), the hysteretic character of the 
laminar-turbulent transition was characterized in 
much detail (Herault {\it et al.} 2015).

\begin{figure}[ht]
\begin{center}
\includegraphics[width=0.85\textwidth]{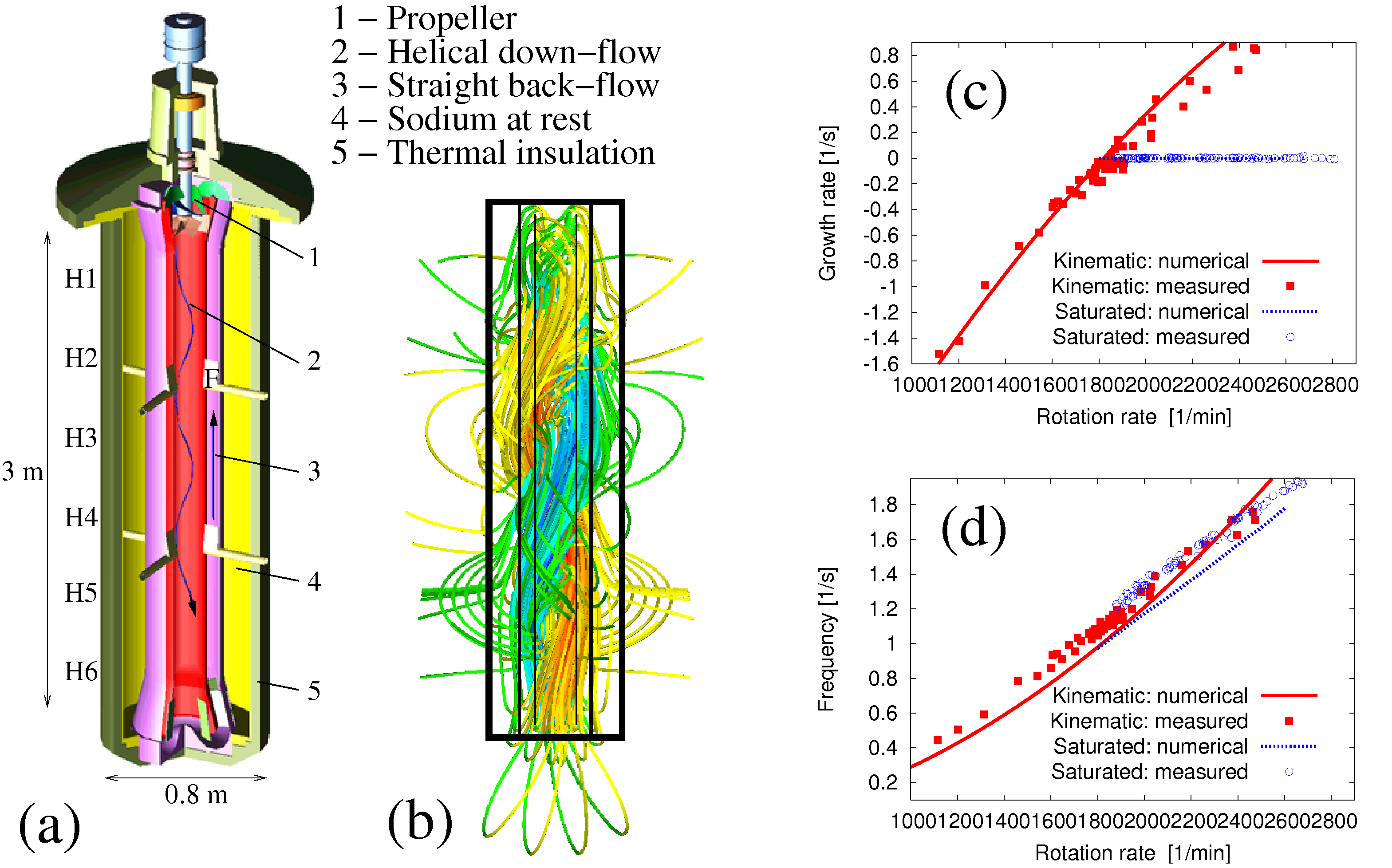}
\caption{The Riga dynamo experiment. (a) Scheme. (b) 
Simulated magnetic eigenfield in form of a double helix 
structure that rotates around the vertical axis with a 
frequency between 1 and 2 Hz. Experimentally determined 
and numerically computed growth rates (c) and frequencies 
(d), both in the kinematic and in the saturated regime 
of the dynamo (after Gailitis {\it et al.} 2009).}
\end{center}
\end{figure}

Based on the theoretical work of Ponomarenko (1973), 
and on further optimization by Gailitis and Freibergs (1976), 
an early version of the later Riga dynamo experiment was 
carried out in 1987 (Gailitis 1987). 
After having 
reached a significant amplification of an applied magnetic 
field, the experiment had to be stopped due to technical 
problems. The very Riga dynamo, whose ultimate success 
relied on the iterative optimization 
of the flow profile (Stefani et al. 1999), 
became operative in November 1999. 
Besides the strong amplification of an externally applied 
field (up to a factor 25), we finally observed a slowly 
growing magnetic eigenfield, i.e. a kinematic dynamo, but 
for only 15 seconds (Gailitis {\it et al.} 2000), 
after which the maximal rotation 
rate had to be reduced for safety reasons. In July 2000, 
the saturated regime of this dynamo was also reached 
(Gailitis {\it et al.} 
2001). 
Since that time, 10 experimental 
campaigns haven been carried out 
(Gailitis {\it et al.} 2018), 
the last ones in June 2016 (Gailitis and Lipsbergs 2017) 
and April 2017, 
with an ever increasing refinement of measurement techniques. 
Figure 1 shows, together with a sketch of the facility and 
a simulation of the magnetic eigenfield, a compilation of 
the main results in form of the experimentally and numerically 
determined growth rates and eigenfrequencies, both in the 
kinematic and in the saturated regime.

Nearly simultaneously with the single-scale 
Riga facility, the two-scale 
dynamo experiment in Karlsruhe became operative and 
immediately provided self-excitation and saturation 
of a magnetic eigenfield 
(M\"uller and Stieglitz  {\it et al.}
2000, Stieglitz and M\"uller 2001, 
M\"uller  {\it et al.} 2004). Similar as in the 
Riga case, the results of this experiment were in nearly 
perfect agreement with numerical predictions 
(R\"adler {\it et al.} 1998, 
Tilgner 2002, R\"adler {\it et al.} 2002). 
More surprising were the results of the von-Kármán-sodium 
(VKS) dynamo experiment in Cadarache that reached the 
threshold of self-excitation in 2006 
(Monchaux {\it et al.} 2007). 
The reasons for the discrepancy between the 
numerically predicted and the experimentally 
observed field topology are still a 
matter of ongoing research: the soft iron impellers 
seem to play a big role here, as discussed in a number of 
numerical simulations  (Giesecke {\it et al.} 
2010, Giesecke {\it et al.} 2012, Nore {\it et al.} 2015,
Kreuzahler {\it et al.} 2017).  This notwithstanding,
the results of the VKS experiment are fascinating, 
in particular with view on its rich dynamical behaviour, 
including field reversals and excursions
(Monchaux {\it et al.} 2009, Miralles {\it et al.} 2013).

By now, dynamo (related) experiments are carried 
out at many places in the world 
(Stefani {\it et al.} 2008, 
Lathrop and Forest 2011). 
Among their 
most important results are the observation of an 
unexpected axi-symmetric induced field at the dynamo 
experiment in Madison (Spence {\it et al.} 2006), 
the accurate determination 
of a turbulence-enhanced resistivity in Perm 
(Frick {\it et al.} 2010), 
the 
identification of an MRI-like mode in a turbulent 
spherical Couette flow in Maryland (Sisan {\it et al.} 2004),
the measurement of a 
strong $\Omega$ - effect in Socorro 
(Colgate {\it et al.} 2011), 
and the 
observation of the numerically predicted 
super-rotation at the rotating sphere experiment 
in Grenoble (Schmitt {\it et al.} 2013). 
The rapidly rotating liquid sodium experiment in Zurich, 
with a radial electrical current and an imposed axial 
magnetic field,  will allow to study 
the magnetostrophic regime 
(Hollerbach {\it et al.} 2013).
And  the old dream of a homopolar disk dynamo
might finally come true in an experiment in Quer\'etaro (Mexico) 
which uses an
optimized system of a  
spirally slotted copper disk and liquid metal 
contacts (Avalos-Zu\~{n}iga {\it et al.} 2017) .

\subsection{Experiments on magnetically triggered 
flow instabilities}

The preparatory decade of the first liquid sodium 
dynamo experiments concurred with the recognition 
of the fundamental importance of the MRI for 
accretion disk physics 
(Balbus and Hawley 1991, Balbus 2003). 
It soon became clear that the demonstration 
of SMRI, with a purely axial field being applied,
would require similar experimental effort as dynamo 
experiments do (R\"udiger {\it et al.} 2003, 
Liu {\it et al.} 2006a). 
A corresponding  experiment was 
constructed in Princeton and has shown  
interesting results on slow magneto-Coriolis 
waves (Nornberg {\it et al.} 2010) and on the magnetic 
destabilization of a free Shercliff layer 
(Roach {\it et al.} 2012), although not yet the very 
SMRI.

\begin{figure}[ht!]
\begin{center}
\includegraphics[width=0.85\textwidth]{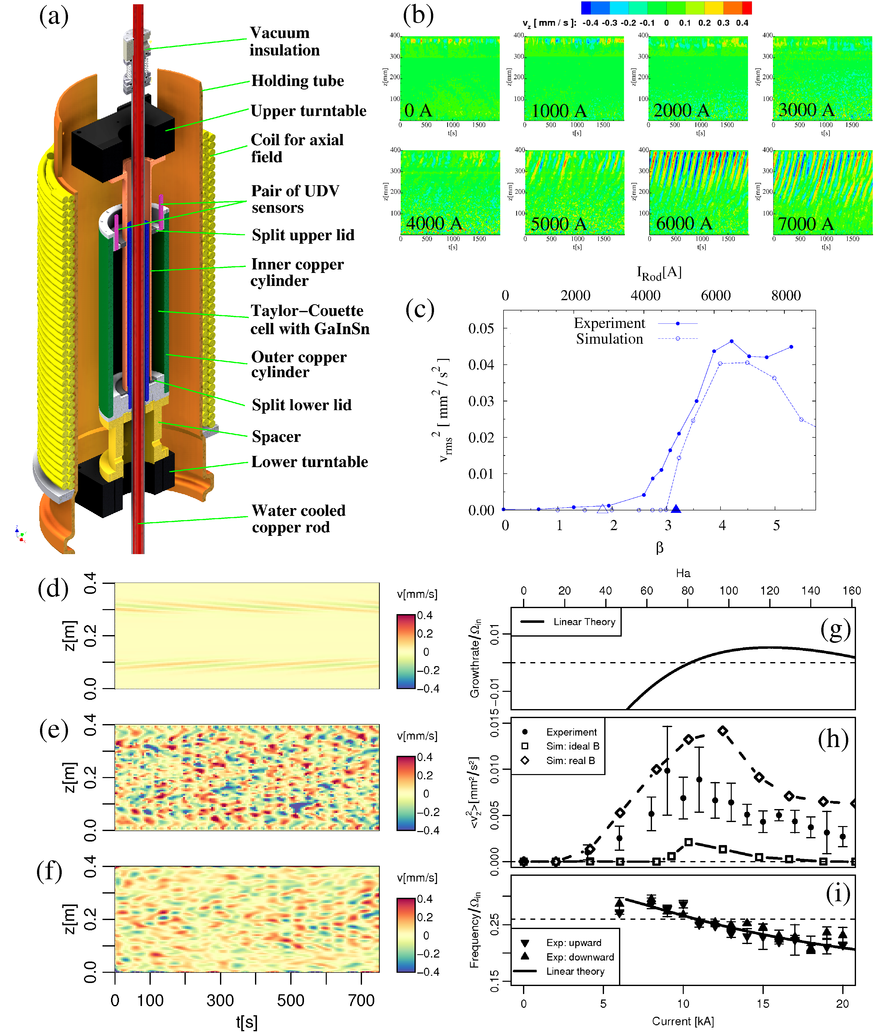}
\caption{The PROMISE experiment at HZDR, and some of 
its main results on HMRI (b,c) and AMRI (d-i). 
(a) Sketch of the facility, comprising a Taylor-Couette 
cell with inner radius  $r_i=4$\,cm, outer radius $r_o=8$\,cm, 
and a height $H = 40$\,cm, filled with GaInSn, an 
external coil and a central copper rod for the 
generation of the axial and the azimuthal magnetic 
field, respectively. (b) Measured velocity structure 
when increasing the current through the central rod, 
for a fixed current of 76 A in the coil. 
Approximately at 5000 A, HMRI emerges as an 
upward travelling wave. (c) Experimentally and 
numerically determined dependence of the mean squared 
of the velocity perturbations on the applied rod 
current. Velocity perturbation $v_z(m=1; z; t)$ 
for $Re = 1480$ and $Ha = 124$ for AMRI (after Seilmayer 
{\it et al.} 2014): (d) 
simulation for an idealized axisymmetric field. 
(e) Simulation for the realistic field geometry. 
(f) Measured velocity. (g) Numerically predicted 
growth rate. (h) Simulated and measured mean 
squared velocity perturbation. (i) Angular drift frequency.}
\end{center}
\end{figure}

Since 2005 an experimental research 
programme has been pursued at HZDR which was devoted 
to the characterization of the 
helical (HMRI) and azimuthal version (AMRI) of 
the MRI in a Taylor-Couette flow with the eutectic 
metal GaInSn, which is liquid at room temperature 
and thus allows much simpler settings than liquid 
sodium experiments. The basic idea behind these 
PROMISE experiments (Stefani {\it et al.} 2006, 
Stefani  {\it et al.} 2009, Seilmayer {\it et al.} 2014) 
is the fact that  HMRI and AMRI work already 
at much lower values of the Reynolds ($Re$) and the 
Hartmann ($Ha$) number than SMRI
(Hollerbach and R\"udiger 2005, 
Hollerbach {\it et al.} 2010). The 
reason for the different scaling behaviour of HMRI 
and SMRI is that SMRI represents a destabilized 
slow magneto-Coriolis wave, while HMRI is 
essentially a weakly destabilized inertial oscillation. 
Nevertheless, both instabilities are connected 
continuously (Hollerbach and R\"udiger 2005), 
a fact that was explained 
in terms of the appearance of an exceptional point 
where both modes coalesce and exchange their branches 
(Kirillov and Stefani 2010).

In a first version of this experiment (PROMISE 1), 
the appearance of the HMRI was already observed 
in the predicted window of $Ha$ 
(Stefani {\it et al.} 2006, Stefani {\it et al.} 2007), 
starting from an axial current of 
around 4 kA.  An imperfection of this first 
version was the strong influence of the Ekman 
pumping at the top and bottom lids 
which resulted in a 
radial jet (approximately at mid-height of the 
Tayler-Couette cell) where the MRI wave ceased 
to exist. To minimize this Ekman pumping 
the experiment was modified by installing end-rings that 
were split at a numerically optimized position (Szklarksi 2007). 
This version, PROMISE 2, is illustrated in 
Figure 2a. As a consequence of this apparently 
slight modification, the HMRI wave travelled 
throughout the entire height (Figure 2b) 
(Stefani {\it et al.} 2009), 
and the coincidence with numerical predictions 
became much better than in PROMISE 1 (see Figure 2c).

A third version, PROMISE 3, which required a significant 
enhancement of the power supply to provide central 
currents of up to 20 kA, 
has confirmed the numerical 
prediction of AMRI (Hollerbach {\it et al.} 2010). 
A peculiarity 
of this experiment was the surprisingly strong effect 
of the slight symmetry breaking of the applied 
field (owing to the one-sided wiring for 
the central current). While a perfectly symmetric 
field would have led to an instability pattern 
according to Figure 2d, the correct consideration 
of the symmetry breaking leads to the 
interpenetrating wave pattern as shown in 
Figure 2e, which was indeed observed in 
the experiment (Figure 2f). With improved 
numerics a reasonable agreement with 
the measured rms values (Figure 2h) and 
frequencies (Figure 2i) of the AMRI wave
was achieved (Seilmayer {\it et al.} 2014). 

Further to HMRI and AMRI, in another GaInSn experiment 
first evidence of the Tayler instability 
(TI) was gained, which 
expectedly set in at a current of around 3 kA 
running axially through a liquid metal column 
(Seilmayer {\it et al.} 2012).

\section{The DRESDYN project}

From the basic research viewpoint, 
the main aim of the DRESDYN project at HZDR 
is to 
contribute
to a better understanding of the origin and the 
action of planetary, stellar, and galactic 
magnetic fields. Dedicated liquid sodium experiments 
will be carried out in parameter regions that 
are hardly accessible to numerics. 
Figure 3 gives some illustrations of the 
present status of the DRESDYN building, 
which is now ready to host 
the experiments.

\begin{figure}[h!]
\begin{center}
\includegraphics[width=0.9\textwidth]{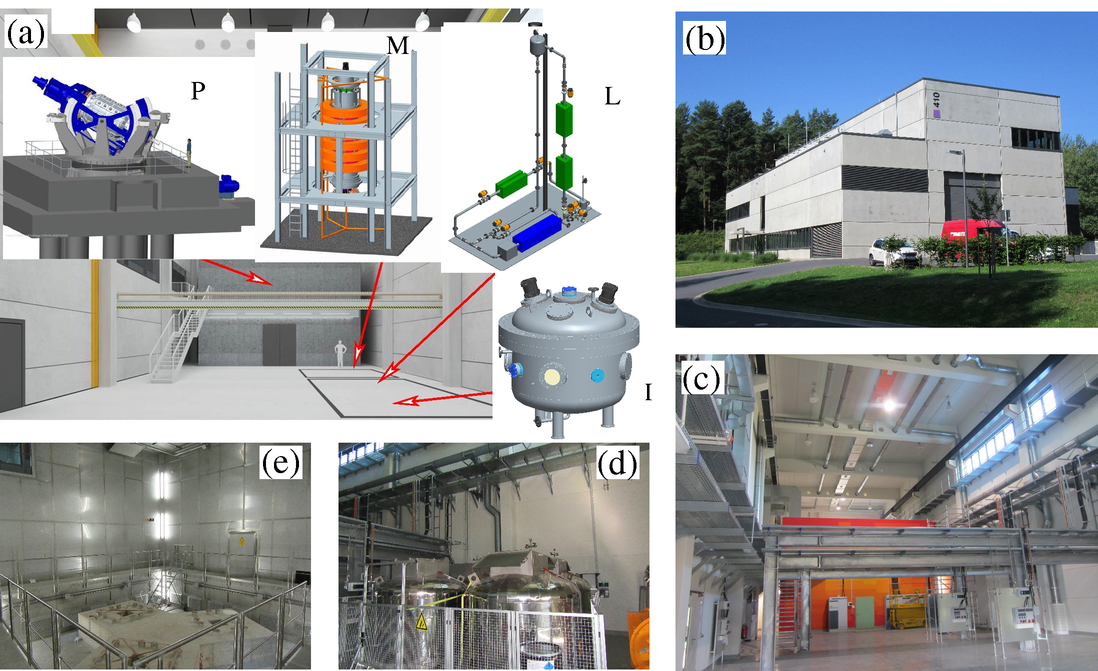}
\caption{Overview about the DRESDYN facility 
and impressions of the building and infrastructure. 
(a) Scheme of the experimental hall with the major 
planned experiments: P - Precession experiment in a 
special containment, M - MRI/TI experiment, 
L - Sodium loop for testing measurement techniques, 
I - Sodium pool facility.  
(b) Finalized building from outside. 
(c) View into the experimental hall. (d) Containers 
for a total of 12 tons of sodium. (e) View into the 
safety containment with the massive ferro-concrete 
tripod for the precession experiment.
}
\end{center}
\end{figure}

The basic studies in the frame of DRESDYN have 
three specific scientific goals: First, 
to demonstrate magnetic 
field self-excitation in a truly homogeneous fluid, 
which is driven in a ''natural'' manner by precession.  
Second, to perform a combined MRI/TI 
experiment, with the particular aim to find the 
"holy grail" of demonstrating standard MRI, and 
third, to set-up and run an experiment on 
''Super-AMRI'', which is supposed to confirm the 
surprising prediction of magnetic 
destabilization of rotating flows with positive shear.
The large dimensions of the precession experiment 
(and, to some lesser extent, of the 
MRI/TI experiment, too)
with several tons of liquid sodium, and its 
operation at the edge of technical feasibility, 
make DRESDYN a somewhat risky endeavor. 

\subsection{Precession dynamo}

A mass of 6 tons of liquid sodium, set 
into motion only by the simultaneous rotation 
around two axes, without using any impellers or guiding 
blades, may indeed be considered {\it the} 
paradigm of a homogenous 
hydromagnetic dynamo. Precession has long been 
discussed as an energy source of the geodynamo 
(Malkus 1968,
Kerswell 1993, Tilgner 2005, Tilgner 2007, 
Wu and Roberts 2009, Shalimov 2006, Nore {\it et al.} 2011, 
Goepfert and Tilgner 2016) and the ancient lunar dynamo 
(Dwyer {\it et al.} 2012, 
Noir and Cebron 2013, 
Weiss and Tikoo 2014), and 
the link between various Milankovic cycles of the 
Earth's orbit parameters and the occurrence of 
reversals (Consolini and De Michelis 2003, 
Fischer {\it et al.} 2009) 
makes it a very interesting 
subject of geophysics in general. 

\begin{figure}[h!]
\begin{center}
\includegraphics[width=0.9\textwidth]{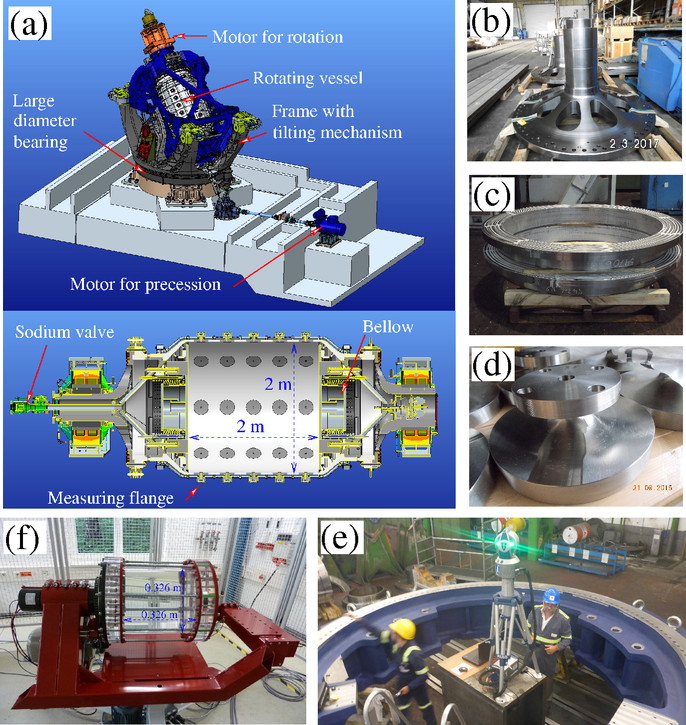}
\caption{Precession experiment: (a) Design of the 
entire machine (top) and the central module (bottom). 
Note that the precession angle can be varied 
between 45° and 90° (figure courtesy SBS B\"uhnentechnik GmbH). 
(b) Photographs of the 
produced shaft flanges. (c) Two rings of the 
container. (d) Measuring flanges. (e) Impression 
from the conformance inspection of the large cast 
ring to be installed between the ferro-concrete 
tripod and the large ball bearing. (f) The 1:6 
down-scaled water experiment for predicting 
the flow and pressure fields of the large experiment.
}
\end{center}
\end{figure}

A few remarks on the homogeneity of dynamo experiments: 
The Riga dynamo 
had already a rather high degree of homogeneity, 
with a length of 3 m along which the flow was 
susceptible to the back-reacting Lorentz forces 
resulting from the self-excited magnetic field. 
Consequently, the saturation curve (i.e., the 
dependence of the magnetic field energy on the 
super-criticality) is quite flat (see Figure 5 
in Gailitis {\it et al.} 2008), which 
reflects the ''evasive'' character of this 
fluid dynamo. Nevertheless, two stainless 
steel tubes were still needed to separate 
the inner helical flow from the back-flow, and the 
back-flow from the outer sodium at rest. A significant 
part of the saturation mechanism relies on the 
decrease of the differential rotation between 
down-flow and back-flow (Gailitis {\it et al.} 2004), 
which is of course strongly influenced by the 
presence of a wall.  The 
Karlsruhe experiment contained, for the good 
reason to show two-scale dynamo action, 52  
stainless-steel ''spin-generators''. 
Consequently, its saturation 
curve turned out to be quite steep (again, see 
Figure 5 in Gailitis {\it et al.} 2004). 
The VKS dynamo has shown an impressive variety of 
dynamical effects such as reversals and excursions 
(Berhanu {\it et al.} 2007, 
Monchaux {\it et al.} 2009), but only when the 
impellers were made of soft iron, which compromises 
significantly its homogeneity. What we are aiming 
at now 
is to build a truly homogenous 
laboratory dynamo that is driven in a near-natural manner. 

Despite the motivational background of planetary 
dynamos working in (near-)spherical geometry, we 
have opted for a cylindrical geometry due to its 
stronger, pressure driven forcing of the fluid, 
for which a wealth of analytical and numerical 
data exist as well. Much has been learned already 
about the flow structure in such a setting: when 
precessing the cylinder very slowly, the initial 
solid-body rotation is superposed 
by a non-axisymmetric ($m = 1$) Kelvin mode. 
Increasing the precession ratio, more modes 
come into play, until suddenly the quasi-laminar 
flow gives way to a completely turbulent regime. 
While the pure hydrodynamics of precession is 
interesting in itself, and will indeed be studied, 
the main focus of the experiment will be on the 
question in which parameter regime dynamo action 
occurs.

\begin{figure}[h!]
\begin{center}
\includegraphics[width=0.9\textwidth]{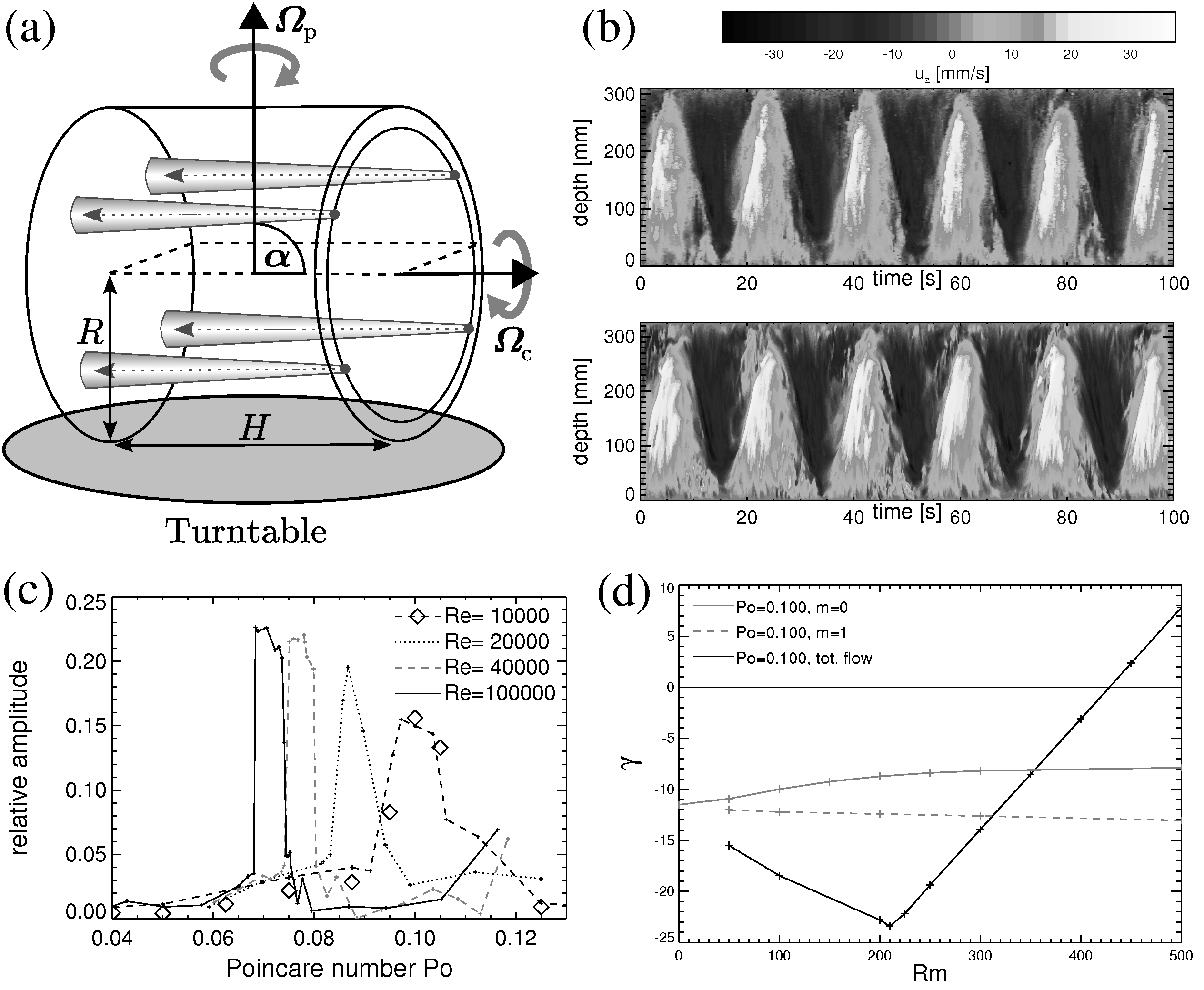}
\caption{The precession driven flow and its predicted 
dynamo action. 
(a) Scheme of the down-scaled water experiment 
(see Figure 4f) with 4 Ultrasonic Doppler Velocity 
(UDV) transducers for flow measurement.  
(b) Temporal evolution of the axial velocity 
$u_z(z,t)$ at a radius $r = 150$ mm (upper panel: UDV measurements, 
lower panel: simulations). 
(c) Relative amplitude of the axisymmetric double-roll 
mode with respect to the dominant $m = 1$ mode for 
various $Re$. The curves denote results from the 
water experiment, and the diamonds denote results 
from simulations. (d) Growth rates of the 
dynamo-generated magnetic field versus $Rm$ for 
combinations of various azimuthal modes from the 
velocity field obtained at simulations at 
$Re = 10^4$ and $Po = 0.1$. The vertical dashed 
lines indicate the respective dynamo thresholds. 
The full flow leads to a critical $Rm = 428$
(after Giesecke {\it et al.} 2018). 
}
\end{center}
\end{figure}

The precession experiment is the most ambitious 
project within the framework of DRESDYN. The design 
of the experiment is finalized (see Figure 4a), 
the construction is ongoing, and first test-experiments 
with water are expected for 2019. The wealth of 
results (Herault {\it et al.} 2015, Giesecke {\it et al.} 2015,
Stefani {\it et al.} 2017,
Giesecke {\it et al.} 2018, Herault {\it et al.} 2018), 
obtained at a 1:6 
down-scaled water-dummy experiment (Figure 4f) 
working up to $Re = 1.5 \times 10^6$, and by numerical 
simulations up to $Re = 10^4$, provides a sound basis 
for the large-scale liquid sodium experiment. 
Since the centrifugal pressure (up to 20 bar), 
in combination with the precession-driven pressure 
swing (up to 10 bar) (Stefani {\it et al.} 2017), 
make the safe construction 
of the cylindrical shell a great challenge, the 
first water experiments will serve for confirming 
the ultimately allowable rotation and precession 
rates which by now are supposed to be 10 Hz and 1 Hz, 
respectively.

As for the dynamo action of the precession-driven 
flow, we have recently obtained a very promising 
result (Giesecke {\it et al.} 2018) which helps 
constraining the parameter 
regime to be explored (Figure 5). This result 
relies on the numerical and experimental finding that, 
shortly before the transition from the laminar to the 
turbulent flow, the non-linear self-interaction of 
the dominant forced $m = 1$ Kelvin mode produces an 
axisymmetric ($m = 0$) mode with doubled axial 
wavenumber (Figure 5c). The resulting double-roll 
structure is known to be an efficient dynamo 
(Dudley and James 1983). 
Indeed, numerical dynamo simulations have shown 
a critical magnetic Reynolds number 
of $Rm_c = 428$ (Figure 5d), which is safely 
below the experimentally achievable value of $Rm = 700$. 
Despite some remaining ambiguities, concerning the 
scaling with the Reynolds number and the detailed 
role of electrical boundary conditions, this 
double-roll mode has been experimentally shown to 
be a robust phenomenon, which seems also be linked to the 
hysteretic behaviour of the laminar/turbulent 
transition as observed previously (Herault {\it et al.} 2015). 
Hence, one might indeed be 
optimistic to find dynamo action close to a 
precession ratio of $Po:=f_{\rm prec}/f_{\rm cyl} \approx 0.06$, 
and at around 2/3 of the 
available rotation speed.
Further to this, it is also planned to explore the turbulent 
regime, for which neither the flow structure nor 
its dynamo action is well understood. While we 
expect a significant amount of small-scale helicity 
to build up in this regime, it could be possible 
that its dynamo capability is thwarted by the simultaneous 
increase of the turbulent resistivity ($\beta$-effect), an 
effect that might have played a detrimental role in 
previous experimental dynamo studies 
(Rahbarnia {\it et al.} 2012).

\subsection{MRI/TI experiment}

The aim of the combined MRI/TI experiment is two-fold: 
First, we want to study in detail the transition between 
the helical MRI (HMRI) and the standard MRI (SMRI), by 
simultaneously increasing the Reynolds and Hartmann 
numbers and decreasing the ratio of azimuthal to axial 
field (Hollerbach and R\"udiger 2005). 
The relation between SMRI 
and HMRI, in particular with respect to the scaling 
behaviour, has spurred a lot of theoretical and numerical 
activities. The continuous transition between them 
relies on the formation of an exceptional point at 
which the slow magneto-Coriolis wave and the inertial 
wave coalesce and exchange their branches (Kirillov and Stefani 2010). 
In order to see 
this transition in experiment we have to go from 
$Re\sim 10^3$ and $Ha \sim 10$ to $Re\sim 10^6$ 
and $Ha\sim 10^3$. The Princeton 
experiment had proved the demonstration of 
SMRI to be extremely challenging for the simple reason 
that purely hydrodynamic flows with $Re\sim 10^6$, when 
axially bounded, are extremely hard to keep stable, 
whatever linear stability analysis for Tayler-Couette 
flows with infinite length tells us. The 
provision of a central current in our experiment 
has the crucial advantage that we can start from 
the well-known regime 
of HMRI, and approach from there the SMRI regime 
in a continuous manner. 

\begin{figure}[h!]
\begin{center}
\includegraphics[width=0.9\textwidth]{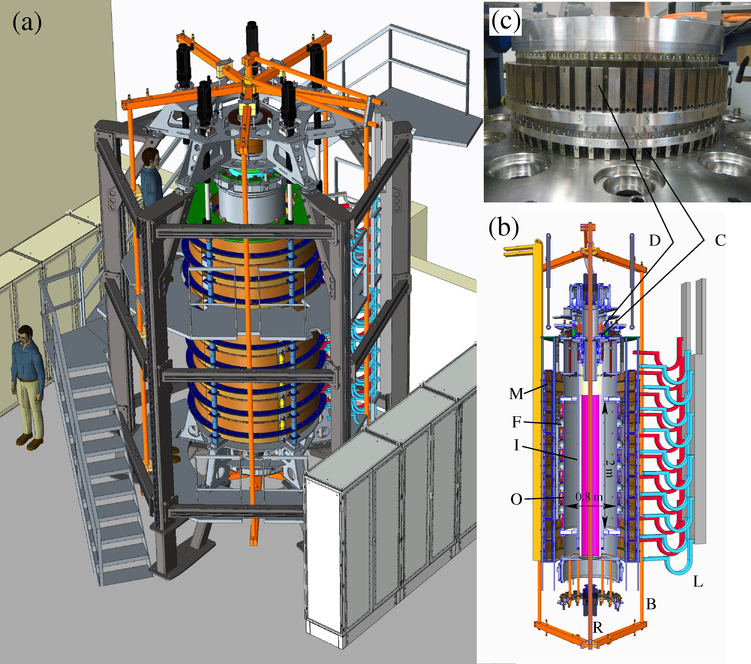}
\caption{The liquid sodium experiment for the combined 
investigation of MRI and TI. (a) Drawing of the entire 
facility. (b) Drawing of the central module: 
I- Inner cylinder, O - Outer cylinder, F - Measurement 
flanges, M - Magnet for the axial 
field, R - Central rod for the azimuthal 
field, B - Back-wires for return current, D - Drive 
for the inner cylinder, C - Magnetic coupler for the 
inner cylinder (figure courtesy S. K\"oppen). 
(c) Photograph of the drive and the 
magnetic coupler.}
\end{center}
\end{figure}

The second goal is to study various combinations 
of AMRI (or HMRI) with TI by guiding additional 
currents through the liquid. As revealed recently
(Kirillov and Stefani 2013, 
Kirillov {\it et al.} 2014, R\"udiger {\it et al.} 2015), 
this combination allows to destabilize 
even Keplerian profiles at very moderate values of 
$Re$ and $Ha$, which was not possible in the HMRI and 
AMRI experiments at PROMISE with its purely 
central current through the hole (Stefani {\it et al.} 2006, 
Seilmayer {\it et al.} 2014).

The construction of the liquid-sodium MRI/TI 
experiment (Figure 6) relies on our experience 
with HMRI and AMRI gained at the PROMISE facility, 
and with the TI experiment. One of the main 
constructional challenges of the new device 
is to combine a 
sophisticated mechanical configuration of split 
end-rings (which is necessary for the MRI part) 
with appropriate electrical contacts for applying 
the internal currents that are needed for the TI. 
As shown by R\"udiger {\it et al.} (2003), 
MRI for liquid sodium starts at a magnetic Reynolds 
number of $Rm=21$ and a Lundquist 
number of $Lu=4.4$ (both values correspond to 
the more conservative estimate for 
different electrical boundary conditions). 
With a height of the active Tayler-Couette cell 
of 2 m (Figure 2b), and an outer radius of  0.4 m, 
we plan to reach a rotation rate of the inner 
cylinder of 20 Hz and an axial field of 130 mT, 
which is in either case more than double the 
critical value. With view on the safety issues 
when dealing with one ton of liquid sodium, 
one of the most critical aspects is the driving 
of the inner cylinder. For that purpose, a 
sophisticated magnetic coupler (Figure 6c) 
has been developed and already tested, which 
ensures a hermetic seal of the entire experiment. 
Another critical part of the experiment is the 
large coil for generating the axial magnetic 
field. Great effort was spent to make this 
field as homogeneous as possible. The resulting 
construction weighs 5 tons, and will require 
around 120 kW of electrical power. 

First experiments are planned for 2020. 
For the sake of its overwhelming scientific 
relevance, we will first try to approach standard 
MRI from the well-known regime of helical MRI. 
The procedure to do so was already indicated 
in the original paper by Hollerbach and Rüdiger 
(2005) where the optimal parameters for transition 
between HMRI and SMRI were determined. 
The study of combinations of AMRI and HMRI 
with TI is planned for a later stage of the 
experiment. In addition to the currents 
through the central rod, for the TI part we will 
also need currents through the liquid. In the 
foreseen geometry with a ratio of inner to 
outer radius of 0.5, the Tayler instability 
will set in at around 1 kA for non-rotating 
cylinders. For rotating cylinders this number 
increases according to R\"udiger and Schulz (2010). 

\subsection{Super-AMRI}

As shown recently, both in the 
framework of a short-wavelength approximation 
(Stefani and Kirillov 2015) and by a 1D-stability code 
(R\"udiger {\it et al.} 2016, 
R\"udiger {\it et al.} 2018a), a 
conducting fluid with very strong positive 
shear might develop an $m=1$ instability (''Super-AMRI'')
when subjected to an azimuthal 
magnetic field. Similar effects occur, for 
optimized ratios of azimuthal to axial field, 
also for other azimuthal wavenumbers, including 
$m=0$ (''Super-HMRI''). This surprising result becomes 
plausible when relating the growth rate 
of HMRI with the non-modal growth factor 
of purely hydrodynamic flows, which leads to 
a very simple and nearly linear relationship 
(Mamatsashvili and Stefani 2016). As a brief astrophysical 
digression: given the rather small magnetic Prandtl 
number of the solar tachocline (10$^{-3}$-10$^{-2}$), 
which is actually not that far from those of liquid 
metals (10$^{-6}$-10$^{-5}$), we expect to gain new 
insight into the dynamics of its equator-near 
parts which exhibit indeed positive shear, with possible 
consequences for our alternative model of the solar dynamo 
(Stefani {\it et al.} 2016b).

\begin{figure}[h!]
\begin{center}
\includegraphics[width=0.9\textwidth]{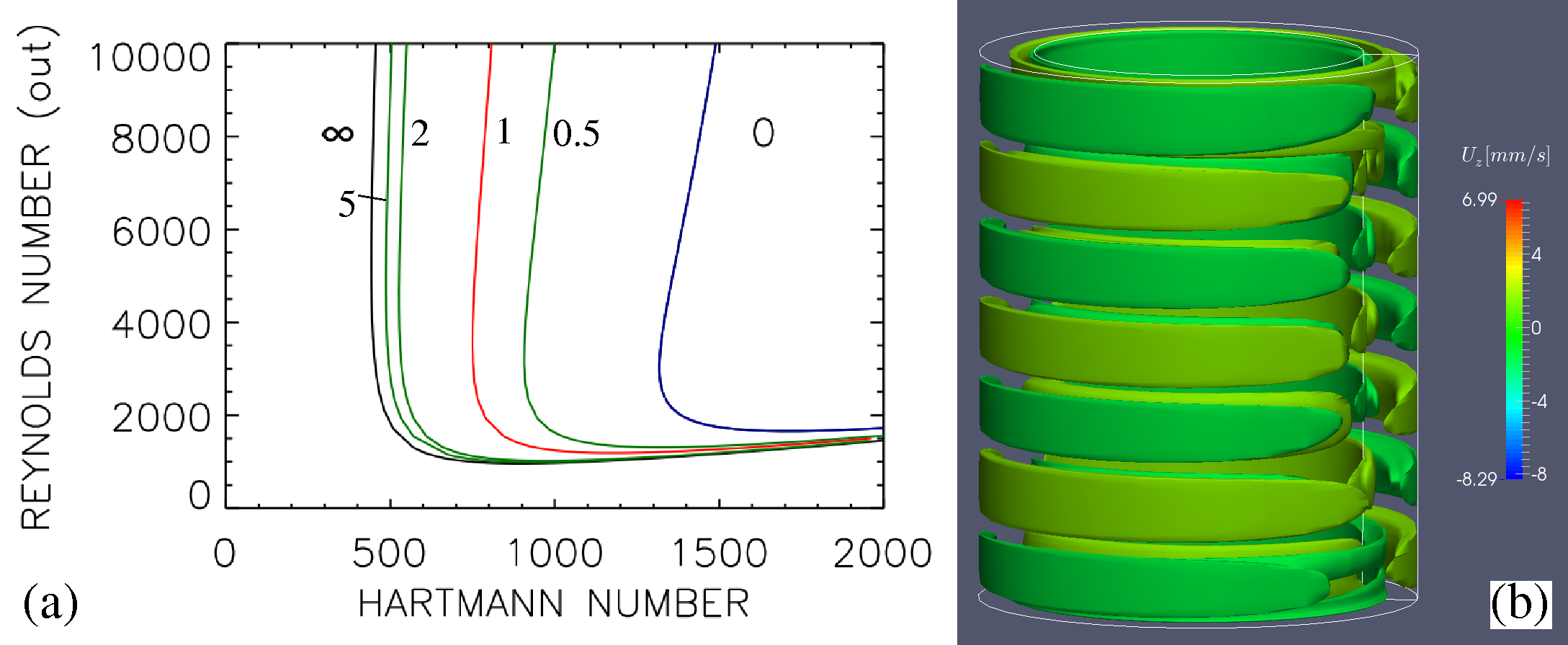}
\caption{Numerical simulations for the Super-AMRI experiment. 
(a) Stability threshold in dependence on $Ha$ and 
$Re$ for a radius ratio $r_i/r_o=0.9$. The numbers 
in the plot indicate the assumed ratio of wall conductivity 
to fluid conductivity. A ratio of 5, which correspond 
approximately to a copper wall for liquid sodium, 
is close to the limit of infinite wall conductivity. 
(b) Simulated 
axial velocity perturbation for $r_i/r_o=0.75$ 
(Figure courtesy to M. Gellert). }
\end{center}
\end{figure}

While the critical Reynolds number for
Super-AMRI is only a 
few thousand, which presents no problem for a 
respective experiment, the predicted critical 
values of the central current range from 
25 to 80 kA, depending strongly on the electrical 
boundary conditions. This is indeed a critical point, 
even under the assumption that a current source of 
50 kA will be available for the combined MRI/TI 
experiment. Very recent numerical 
simulations,  taking into account 
realistic combinations of the conductivities 
of the fluid (sodium) and the wall (copper, 
for example), had revealed a technical 
optimum of around 33 kA for a ratio of inner 
to outer cylinder
radii of $r_i/r_o=0.78$ (R\"udiger {\it et al.} 2018b).

A corresponding Super-AMRI experiment 
is still in its conceptual phase. 
Assuming the radii ratio to be as 
small as 0.75, as in Figure 7b, then a certain 
minimum of the Taylor-Couette gap will be 
approximately 3 cm, which is needed for the 
UDV sensors to distinguish between up- and 
down-flowing fluid parcels. This value would 
translate to a minimum radius of the experiment 
of 12 cm. Further optimization will definitely 
be needed
here. It remains also to be tested whether
the central current can still be reduced for 
a corresponding Super-HMRI experiment.

\section{Conclusions and prospects}

In this short review, we have delineated the 
progress in the experimental investigation of 
the dynamo effect and of magnetically triggered flow
instabilities that was gained during the last two decades. 
With the DRESDYN project, we plan to 
continue this research programme by showing dynamo
action in a homogenous fluid driven by precession, 
and by approaching standard MRI in a Tayler-Couette
experiment. Additional experiments on the magnetic 
destabilization of positive shear flows are also 
planned.
While having focused on liquid metal experiments, we 
see also a huge potential for corresponding plasma 
experiments which have the great advantage of 
allowing to adjust the magnetic Prandtl number,
and whose scientific exploitation  has just begun 
(Forest {\it et al.}  2015, 
Weisberg {\it et al.} 2017).

\section*{Acknowledgments}
We thank Bernd Wustmann, 
Christian Steglich, Frank Herbrand, 
and Sebastian  K\"oppen for their technical assistance 
in the design and construction of various 
experiments. The long-standing collaboration with 
Marcus Gellert, Rainer Hollerbach, Oleg Kirillov, 
George Mamatsashvili and Manfred Schultz  is 
gratefully acknowledged.

\section*{References}

\noindent
Avalos-Zu\~{n}iga, R.A., Priede, J.  and Bello-Morales, C.E., 
A
homopolar disk dynamo experiment with liquid metal contacts. 
\textit {Magnetohydrodynamics} 2017, \textbf{53}, 
341-347.\\[0.5mm]

\noindent
Balbus, S.A., Enhanced angular momentum transport 
in accretion disks. 
\textit {Ann. Rev. Astron. Astrophys.} 2003, \textbf{41}, 
555-597.\\[0.5mm]

\noindent
Balbus, S.A. and Hawley, J.F., A powerful local shear instability 
in weakly magnetized disks. 1. Linear Analysis. 
\textit {Astrophys. J.} 1991, \textbf{376}, 
214-221.\\[0.5mm]

\noindent
Berhanu, M. {\it et al.}, 
Magnetic field reversals in an experimental turbulent dynamo.
\textit{EPL} 2007, \textbf{77}, 59001.\\[0.5mm]

\noindent
Consolini, G. and De Michelis, P.,
Stochastic resonance in geomagnetic polarity reversals.
\textit{Phys. Rev. Lett.} 2003, \textbf{90}, 058501.\\[0.5mm]

\noindent
Colgate, S.A. {\it et al.}, 
High magnetic shear gain in a liquid 
sodium stable Couette flow experiment: A prelude to an $\alpha-\Omega$ 
dynamo.
\textit{Phys. Rev. Lett.} 2011, \textbf{106}, 175003\\[0.5mm]

\noindent
Dudley, M.L. and James, R.W., 
Time-dependent kinematic dynamos with stationary flows.
\textit{Proc. R. Soc. A} 1989, \textbf{425}, 407-429.\\[0.5mm]

\noindent 
Dwyer, C.A., Stevenson, D.J. and Nimmo, F.,
A long-lived lunar dynamo driven by continuous mechanical stirring.
\textit{Nature} 2012, \textbf{220} 47-61.\\[0.5mm]

\noindent
Fischer, M., Gerbeth, G., Giesecke, A. and Stefani, F.,
Inferring basic parameters of the geodynamo from sequences 
of polarity reversals.
\textit{Inverse Probl.} 2009, \textbf{25} 065011.\\[0.5mm]

\noindent 
Forest, C.B. {\it et al.},
The Wisconsin plasma astrophysics laboratory.
\textit{J. Plasma Phys.} 2015, \textbf{81}, 345810501.\\[0.5mm]

\noindent
Frick, P. {\it et al.}, 
Direct measurement of effective magnetic 
diffusivity in turbulent flow of liquid sodium.
\textit{Phys. Rev. Lett.} 2010, \textbf{105}, 184502\\[0.5mm]

\noindent
Fu, R.R. {\it et al.}, An ancient core dynamo in asteroid Vesta.
\textit{Science} 2013, \textbf{338}, 238-241.\\[0.5mm]

\noindent
Gailitis, A.,  
Theory of the Herzenberg dynamo.
\textit{Magnetohydrodynamics} 1973, \textbf{9}, 445--449.\\[0.5mm]

\noindent
Gailitis, A. and Freibergs, Ya., 
Theory of a helical MHD dynamo.
\textit{Magnetohydrodynamics} 1976, \textbf{12}, 127--129.\\[0.5mm]

\noindent
Gailitis, A. {\it et al.}, Experiment with a liquid-metal 
model of an MHD dynamo. \textit{Magnetohydrodynamics} 1987, 
\textbf{23}, 349-353.\\[0.5mm]

\noindent Gailitis, A. {\it et al.},
Detection of a flow induced magnetic field eigenmode 
in the Riga dynamo facility.
\textit{Phys. Rev. Lett.} 2000, \textbf{84}, 4365-4369.\\[0.5mm] 

\noindent
Gailitis, A. {\it et al.},
Magnetic field saturation in the Riga dynamo experiment.
\textit{Phys. Rev. Lett.} 2001, \textbf{86}, 3024--3027.\\[0.5mm]

\noindent
Gailitis, A., Lielausis, O., Platacis, E.,
Gerbeth, G. and Stefani, F., 
Laboratory experiments on hydromagnetic dynamos.
\textit{Rev. Mod. Phys.}  2002, \textbf{74},  973-990.\\[0.5mm]

\noindent
Gailitis, A., Lielausis, O., Platacis, E.,
Gerbeth, G. and Stefani, F., 
The Riga dynamo experiment.
\textit{Surv. Geopyhs.} 2003, \textbf{24}, 247-267.\\[0.5mm]

\noindent
Gailitis, A., Lielausis, O., Platacis, E.,
Gerbeth, G. and Stefani, F.,
Riga dynamo experiment and its theoretical background.
\textit{Phys. Plasmas} 2004, \textbf{11}, 2838-2843.\\[0.5mm]

\noindent 
Gailitis, A., Gerbeth, G.,  Gundrum, Th., 
Lielausis, O., Platacis, E. and Stefani, F.,
History and results of the Riga dynamo experiments.
\textit{C.R. Phys.} 2008, \textbf{9}, 721-728. \\[0.5mm]

\noindent
Gailitis, A. and Lipsbergs, G., 
2016 year experiments at Riga dynamo facility,
\textit{Magnetohydrodynamics}  2017, \textbf{53}, 349-356.\\[0.5mm]

\noindent
Gailitis, A., Gerbeth, G., Gundrum, Th., Lielausis, O., Lipsbergs, G.,
Platacis, E. and Stefani, F., 
Self-excitation in a helical liquid metal 
flow: The Riga dynamo experiments.
\textit{J. Plasma Phys.}  2018, submitted;  arXiv:1801.01749.\\[0.5mm]

\noindent
Gans, R.F., On hydromagnetic precession in a cylinder.
\textit{J. Fluid Mech.} 1970, \textbf{45}, 111--130.\\[0.5mm]

\noindent Giesecke, A., Stefani, F. and Gerbeth, G.,
Role of soft-iron impellers on the mode selection in the von 
K\'arm\'an-sodium dynamo experiment.
\textit{Phys. Rev. Lett.} 2010, \textbf{104}, 044503.\\[0.5mm] 

\noindent
Giesecke, A. {\it et al.}, 
Influence of high-permeability discs in an 
axisymmetric model of the Cadarache dynamo experiment.
\textit{New J. Phys.} 2012,  \textbf{14}, 053005.\\[0.5mm]

\noindent
Giesecke, A. {\it et al.}, 
Triadic resonances in nonlinear 
simulations of a fluid flow in a precessing cylinder.
\textit{New J. Phys.} 2015, \textbf{17}, 113044.\\[0.5mm]

\noindent
Giesecke, A., Vogt, T., Gundrum, T. and Stefani, F.,
Nonlinear large scale flow in a precessing cylinder 
and its ability to drive dynamo action.
\textit{Phys. Rev. Lett.}, 2018 \textbf{120}, 024502.\\[0.5mm]  

\noindent
Goepfert, O. and Tilgner, A.,
Dynamos in precessing cubes.
\textit{New J. Phys.} 2016, \textbf{18}, 103019.\\[0.5mm]

\noindent
Goto, S., Shimizu, M. and Kawahara, G., 
Turbulent mixing in a precessing sphere.
\textit{Phys. Fluids} 2014, \textbf{26}, 115106.\\[0.5mm]

\noindent
Herault, J., Gundrum, T., Giesecke, A. and Stefani, F.,
Subcritical transition to turbulence of a 
precessing flow in a cylindrical vessel.
\textit{Phys. Fluids} 2015,  \textbf{27}, 124102.\\[0.5mm]

\noindent
Herault, J., Giesecke, A., Gundrum, T. and Stefani, F.,
Instability of precession driven Kelvin modes: evidence of 
a detuning effect.
\textit{Phys. Rev. F} 2018,  submitted.\\[0.5mm]

\noindent
Herzenberg, A., Geomagnetic dynamos. 
\textit{Philos. Trans. R. Soc. Lond.} 1958, \textbf{A250}, 
543--585.\\[0.5mm]

\noindent
Hollerbach, R. and R\"udiger, G.,
New type of magnetorotational instability in cylindrical 
Taylor-Couette flow.
\textit{Phys. Rev. Lett.} 2005, \textbf{95}, 124501.\\[0.5mm]

\noindent
Hollerbach, R., Teeluck, V. and R\"udiger, G., 
Nonaxisymmetric magnetorotational instabilities in 
cylindrical Taylor-Couette flow.
\textit{Phys. Rev. Lett.} 2010, \textbf{104}, 044502.\\[0.5mm]

\noindent
Hollerbach, R. {\it et al.}, 
Electromagnetically driven zonal flows in a 
rapidly rotating spherical shell.
\textit{J. Fluid Mech.} 2013, \textbf{725}, 428-445.\\[0.5mm]

\noindent
Kerswell, R.R., The instability of precessing flow.
\textit{Geophys. Astrophys. Fluid Dyn.} 1993, \textbf{72}, 
107-144.\\[0.5mm]

\noindent
Kirillov, O.N. and Stefani, F., 
On the relation of standard and helical magnetorotational 
instability. \textit{Astrophys. J.} 2010, \textbf{712}, 52-68.\\[0.5mm]

\noindent
Kirillov, O.N. and Stefani, F., 
Extending the range of inductionless magnetorotational
instability. \textit{Phys. Rev. Lett.} 2013, \textbf{111}, 061103.\\[0.5mm]

\noindent
Kirillov, O.N., Stefani, F. and Fukumoto, Y., 
Local instabilities in magnetized rotational flows: 
a short-wavelength approach. 
\textit{J. Fluid Mech.} 2014, \textbf{760}, 591-633.\\[0.5mm]

\noindent
Kreuzahler, S., Ponty, Y., Plihon, N., Homann, H. and Grauer, R.,
Dynamo enhancement and mode selection 
triggered by high magnetic permeability.
\textit{Phys. Rev. Lett.} 2017,  \textbf{119}, 234501.\\[0.5mm]

\noindent
Lagrange, R., Eloy, C., Nadal, F. and Meunier, P., 
Instability of a fluid inside a precessing cylinder.
\textit{Phys. Fluids} 2008,  \textbf{20}, 081701.\\[0.5mm]

\noindent
Lathrop, D.P. and Forest, C.B., Magnetic dynamos in the
lab.
\textit{Phys. Today} 2011,  \textbf{64}(7), 40-45.\\[0.5mm]

\noindent
Lin, Y., Marti, P. and Noir, J., 
Shear-driven parametric instability in a precessing sphere.
\textit{Phys. Fluids} 2015, \textbf{27}, 046601.\\[0.5mm]

\noindent
Liu, W., Goodman, J. and Ji, H.,  
Simulations of magnetorotational instability in a 
magnetized Couette flow.
\textit{Astrophys. J.} 2006a, \textbf{643}, 306-317.\\[0.5mm]

\noindent
Liu, W., Goodman, J., Herron, I. and Ji, H.,  
Helical magnetorotational instability in 
magnetized Taylor-Couette flow. 
\textit{Phys. Rev. E} 2006b, \textbf{643}, 056302.\\[0.5mm]

\noindent
Lowes, F.J. and Wilkinson, I.,
Geomagnetic dynamo - A laboratory model.
\textit{Nature} 1963, \textbf{198}, 1158-1160.\\[0.5mm]

\noindent
Lowes, F.J. and Wilkinso, I.,
Geomagnetic dynamo - An improved laboratory model.
\textit{Nature} 1963, \textbf{219}, 717-718.\\[0.5mm]

\noindent
Malkus, W.V.R.,
Precession of Earth as cause of geomagnetism.
\textit{Science} 1968,  \textbf{160}, 259-264.\\[0.5mm]

\noindent
Mamatsashvili, G. and Stefani, F.,
Linking dissipation-induced instabilities 
with nonmodal growth: The case of helical 
magnetorotational instability.
\textit{Phys. Rev. E} 2016,  \textbf{94}, 051203.\\[0.5mm]

\noindent
Miralles, S.  {\it et al.}, 
Dynamo threshold detection in the von K\'arm\'an sodium experiment.
\textit{Phys. Rev. Lett.} 2013, \textbf{88}, 013992.\\[0.5mm]

\noindent
Monchaux, R.  {\it et al.}, 
Generation of a magnetic field by dynamo action in a 
turbulent flow of liquid sodium.
\textit{Phys. Rev. Lett.} 2007, \textbf{98}, 044502.\\[0.5mm]

\noindent Monchaux, R. {\it et al.}, The 
von K\'arm\'an sodium experiment: 
Turbulent dynamical dynamos.
\textit{Phys. Fluids} 2009, \textbf{21}, 035108.\\[0.5mm]

\noindent
Mouhali, W., Lehner, T., L\'eorat, J. and Vitry, R. 
Evidence for a cyclonic regime in a precessing 
cylindrical container.
\textit{Exp. Fluids} 2012, \textbf{53}, 1693-1700.\\[0.5mm]

\noindent
M\"uller and Stieglitz, R., 
Can the Earth's magnetic field be simulated in the laboratory?
\textit{Naturwissenschaften} 2000,  \textbf{87}, 381-390.\\[0.5mm]

\noindent 
M\"uller, U. and  Stieglitz, R.,
The Karlsruhe dynamo experiment.
\textit{Nonl. Proc. Geophys.} 2002, \textbf{9}, 165-170.\\[0.5mm]

\noindent
M\"uller, U.,  Stieglitz, R. and Horanyi, S., 
A two-scale hydromagnetic dynamo experiment.
\textit{J. Fluid Mech}. 2004, \textbf{498}, 31-71.\\[0.5mm]

\noindent Noir, J. and Cebron, D, 
Precession-driven flows in non-axisymmetric ellipsoids.
\textit{J. Fluid Mech.} 2013, \textbf{737}, 412-439.\\[0.5mm]

\noindent Nore, C., L\'eorat, J., Guermond, J.-L. and
Luddens, F., Nonlinear dynamo action in a precessing cylindrical container.
\textit{Phys. Rev. E} 2011, \textbf{84}, 016317.\\[0.5mm]

\noindent Nore, C., Quiroz, D.C., Cappanera, L. and 
Guermond, J.-L., 
Direct numerical simulation of the axial 
dipolar dynamo in the von K\'arm\'an Sodium experiment.
\textit{EPL} 2016, \textbf{14}, 65002.\\[0.5mm]

\noindent
Nornberg, M.D., Ji, H., Schartman, E., Roach, A.,  
and Goodman, J., 
Observation of magnetocoriolis waves in a 
liquid metal Taylor-Couette experiment.
\textit{Phys. Rev. Lett.} 2010,  \textbf{104}, 074501.\\[0.5mm]

\noindent
Ponomarenko, Y.B., On the theory of 
hydromagnetic dynamos,
\textit{Zh. Prikl. Mekh. \& Tekh. Fiz. (USSR)} 1973,  
\textbf{6}, 47-51.\\[0.5mm]

\noindent R\"{a}dler, K.-H., Apstein, E., Rheinhardt, M. and Sch\"{u}ler,
M., The Karlsruhe dynamo experiment. A mean field approach. \textit{Stud.
Geophys. Geodaet.} 1998, \textbf{42}, 224-231. \\[0.5mm]

\noindent R\"{a}dler, K.-H., Rheinhardt, M., Apstein, E. and Fuchs,
H., On the mean-field theory of the Karlsruhe dynamo experiment. 
\textit{Nonlin. Proc. Geophys.} 2002, \textbf{9}, 171-187.\\[0.5mm]

\noindent
Rahbarnia, K. {\it et al.}, 
Direct observation of the turbulent emf and transport of magnetic
field in a liquid sodium experiment.
\textit{Astrophys. J.} 2012, \textbf{759}, 80.\\[0.5mm]

\noindent
Roach, A. {\it et al.}, 
Observation of a free-Shercliff-layer instability in 
cylindrical geometry.
\textit{Phys. Rev. Lett.} 2012,  \textbf{108}, 154502.\\[0.5mm]

\noindent
 R\"udiger, G., Schultz, M and Shalybkov, D.,
Linear magnetohydrodynamic Taylor-Couette instability for liquid sodium.
\textit{Phys. Rev. E} 2003,  \textbf{67} 046312.\\[0.5mm]

\noindent
R\"udiger, G., Hollerbach, R., and Kitchatinov, L.L.,
\textit{Magnetic processes in astrophysics: theory, 
simulations, experiments}, 2013
(Weinheim: WILEY-VCH).\\[0.5mm]

\noindent
R\"udiger, G., Schultz, M.,  Stefani, F. and Mond, M., 
Diffusive magnetohydrodynamic instabilities beyond the 
Chandrasekhar theorem.
\textit{Astrophys. J.} 2015, \textbf{811}, 84.\\[0.5mm]

\noindent
R\"udiger, G., Schultz, M., Gellert, M. and Stefani, F., 
Subcritical excitation of the current-driven Tayler 
instability by super-rotation.
\textit{Phys. Fluids} 2016, \textbf{28}, 014105.\\[0.5mm] 

\noindent
R\"udiger, G., Schultz, M., Gellert, M. and Stefani, F., 
Azimuthal magnetorotational instability with super-rotation.
\textit{J. Plasma Phys.} 2018, \textbf{84}, 735840101.\\[0.5mm] 

\noindent 
R\"udiger, G., Schultz, M., Stefani, F. and Hollerbach, R.,
Magnetorotational instability in Taylor-Couette flows
between cylinders with finite electrical conductivity.
\textit{Geophys. Astrophys. Fluid Dyn.} 2018, submitted.\\[0.5mm]

\noindent
Schmitt, D. {\it et al.}, 
Magneto-Coriolis waves in a spherical Couette flow experiment.
\textit{Eur. J. Mech. - B/Fluids} 2013,  \textbf{37}, 10-22.\\[0.5mm]

\noindent
Seilmayer, M. {\it et al.}, Experimental evidence for a transient 
Tayler instability in a cylindrical liquid-metal column.
\textit{Phys. Rev. Lett.} 2012, \textbf{108}, 244501.\\[0.5mm]

\noindent
Seilmayer, M. {\it et al.},
Experimental evidence for nonaxisymmetric 
magnetorotational instability in a rotating 
liquid metal exposed to an azimuthal magnetic field.
{\textit Phys. Rev. Lett.} 2014, \textbf{113}, 024505.\\[0.5mm]

\noindent
Sisan, D.R. {\it et al.},
Experimental observation and characterization of 
the magnetorotational instability.
\textit{Phys. Rev. Lett.} 2004, \textbf{93}, 114502.\\[0.5mm]

\noindent
Shalimov, S.L.,
On effect of precession-induced flows in the 
liquid core for early Earth's history. 
\textit{Nonl. Proc. Geophys.} 2006, \textbf{13}, 525-529.\\[0.5mm]

\noindent
Spence, E.J. {\it et al.}, 
Observation of a turbulence-induced large scale magnetic field.
\textit{Phys. Rev. Lett.} 2006, \textbf{96}, 055002.\\[0.5mm]

\noindent
Spruit, H.C., 
Dynamo action by differential rotation in a 
stably stratified stellar interior.
\textit{Astron. Astrophys.} 2002, \textbf{381}, 923-932.\\[0.5mm]

\noindent
Steenbeck, M. {\it et al.}, 
Der experimentelle Nachweis einer 
elektromotorischen Kraft l\"angs eines \"au\ss eren 
Magnetfeldes, induziert durch eine Str\"omung fl\"ussigen Metalls 
($\alpha$-Effekt).
\textit{Mber. Dtsch. Akad. Wiss. Berl.} 1967, \textbf{9}, 714-719.\\[0.5mm]

\noindent Stefani, F., Gerbeth, G.  and Gailitis, A.,
Velocity profile optimization for the Riga dynamo experiment.
In \textit{Transfer Phenomena in Magnetohydrodynamic and electroconducting Flows},
edited by A. Alemany, Ph. Marty and J.-P. Thibault, pp. 31-44, 1999 (Kluwer: Dordrecht).\\[0.5mm]

\noindent
Stefani, F. {\it et al.},
Experimental evidence for magnetorotational 
instability in a Taylor-Couette flow under 
the influence of a helical magnetic field.
\textit{Phys. Rev. Lett.} 2006, \textbf{97}, 184502.\\[0.5mm]

\noindent
Stefani, F. {\it et al.},
Experiments on the magnetorotational 
instability in helical magnetic fields.
\textit{New J. Phys.} 2007, \textbf{9}, 295.\\[0.5mm]

\noindent Stefani, F., Gailitis, A. and Gerbeth, G., 
Magnetohydrodynamic experiments on cosmic magnetic fields. 
\textit{Zeitschr. Angew. Math. Mech.}
2008, \textbf{88}, 930-954.\\[0.5mm]

\noindent
Stefani, F. {\it et al.}, 
Helical magnetorotational instability 
in a Taylor-Couette flow with strongly reduced Ekman pumping.
\textit{Phys. Rev. E} 2009,  \textbf{80}, 066303.\\[0.5mm]

\noindent
Stefani, F. {\it et al.}, 
DRESDYN - A new facility for MHD experiments with liquid sodium.
\textit{Magnetohydrodynamics} 2012, \textbf{48}, 103-114.\\[0.5mm]

\noindent
Stefani, F.  and Kirillov, O.N., 
Destabilization of rotating flows 
with positive shear by azimuthal magnetic fields.
\textit{Phys. Rev. E} 2015, \textbf{92}, 051001(R).\\[0.5mm]

\noindent
Stefani, F. {\it et al.}, 
Towards a precession driven dynamo experiment. 
\textit{Magnetohydrodynamics} 2015, \textbf{51}, 275-284.\\[0.5mm]

\noindent
Stefani, F. {\it et al.}, 
Magnetohydrodynamic effects in liquid metal batteries.
\textit{IOP Conf. Ser.: Mater. Sci. Eng.} 2016a, 
\textbf{143}, 012024.\\[0.5mm]

\noindent
Stefani, F., Giesecke, A., Weber, N. and Weier, T.
Synchronized helicity oscillations: A link between 
planetary tides and the solar cycle? 
\textit{Solar Phys.} 2016b, \textbf{291}, 2197-2212.\\[0.5mm]

\noindent Stieglitz R. and M\"{u}ller U., Experimental demonstration
of a homogeneous two-scale dynamo. \textit{Phys. Fluids} 2001,
\textbf{13}, 561-564. \\[0.5mm]

\noindent 
Szklarski, J., Reduction of boundary effects in the spiral 
MRI experiment PROMISE. \textit{Astron. Nachr.} 
2007, \textbf{328}, 499-506.\\[0.5mm]

\noindent 
Tayler, R.J., Adiabatic stability of stars containing
magnetic fields. 1. Toroidal fields.
\textit{Mon. Not. R. Astron. Soc.} 
1973, \textbf{161}, 365-380.\\[0.5mm]

\noindent
Tilgner, A.
Numerical simulation of the onset of 
dynamo action in an experimental two-scale dynamo.
\textit{Phys. Fluids} 2002, \textbf{14}, 4092-4094.\\[0.5mm]

\noindent
Tilgner, A., 
Precession driven dynamos.
\textit{Phys. Fluids} 2005, \textbf{17}, 034104.\\[0.5mm]

\noindent
Tilgner, A.,
Kinematic dynamos with precession driven flow in a sphere.
\textit{Geophys. Astrophys. Fluid Dyn.} 2007, \textbf{101}, 1-9.\\[0.5mm]

\noindent
Vanyo, P. and Dunn, J.R.,  
Core precession: flow structures and energy.
\textit{Geophys. J.  Int.} 2000, \textbf{142}, 409-425.\\[0.5mm]

\noindent
Velikhov, E.P,
Stability of an ideally conducting liquid flowing between cylinders rotating in a 
magnetic field.
\textit{Sov. Phys. JETP} 1959, \textbf{36}, 995-998.\\[0.5mm]

\noindent
Weisberg, D.B. {\it et al.}, 
Driving large magnetic Reynolds number 
flow in highly ionized, unmagnetized plasmas.
\textit{Phys. Plasmas} 2017, \textbf{24},  056502.\\[0.5mm]

\noindent
Weiss, B.P. and Tikoo, S.M., 
The lunar dynamo.
\textit{Science} 2014, \textbf{346}, 1246753.\\[0.5mm]

\noindent
Wilkinson, I.,
The contribution of laboratory dynamo experiments to our
understanding of the mechanism of generation of planetary magnetic fields.
\textit{Geophys. Surveys} 1984,  \textbf{7}, 107-122.\\[0.5mm]

\noindent
Wondrak, T. {\it et al.},
Contactless inductive flow tomography for a model of continuous steel casting.
\textit{Meas. Sci. Techn.} 2010,  \textbf{21}, 045402.\\[0.5mm]

\noindent
Wu, C.C. and Roberts, P.H.,
On a dynamo driven by topographic precession.
\textit{Geophys. Astrophys. Fluid Dyn.} 2009,  \textbf{103}, 
467-501.\\[0.5mm]

\noindent
Zimmermann, D.S., Triana, S.A., Nataf, H.-C. and Lathrop, D.P., 
A turbulent, high magnetic Reynolds number experimental model 
of Earth's core. 
\textit{J. Geophys. Res. - Sol. Earth} 2010, \textbf{119}, 
4538--4557\\[0.5mm]

\end{document}